\documentclass[prb,twocolumn,showpacs]{revtex4}

\usepackage{graphicx}
\usepackage{epstopdf}%
\usepackage{amsmath,amssymb,dsfont,upgreek}

\usepackage[colorlinks=true, linkcolor=blue, citecolor=blue, urlcolor=blue]{hyperref}

\usepackage{natbib}   %pdfstartview=FitV,

\begin{document}
%\date{June 10, 2008}

\title{Renormalization group study of  a two-valley system with spin-splitting}

\author{Alexander Punnoose}
\affiliation{Physics Department, City College of the City University of New York, New York, NY 10031, USA}
\email{punnoose@sci.ccny.cuny.edu}

\begin{abstract}
Renormalization group equations in a two-valley system with valley-splitting and intervalley scattering are derived in the presence of spin-splitting induced by a parallel magnetic field. The relevant amplitudes in different regimes set by the relative strengths of the spin and valley splittings and the intervalley scattering rate are identified. The range of applicability of the standard formula for the magnetoconductance  is discussed.   \end{abstract}


\maketitle

%\section{Introduction}

%%%%%%%%%%%%
%\vspace{-2\baselineskip}
\section{Introduction}
In two dimensions, an in-plane magnetic field, $B_\parallel$, conveniently couples only to the spin-degrees of freedom leading to spin-splitting of the electron bands. Electron-electron (\textit{e-e}) interactions between the different spin-bands gives rise to a finite magnetoconductance, $\sigma(B_\parallel,T)$,  and hence measurement of $\sigma(B_\parallel,T)$ provides a simple and accurate way of determining the effective spin-related interaction strength~\cite{punnoose07, pudalov_RG_JETP}.  In a disordered two-dimensional electron gas (2DEG), the transport properties at low temperatures, $k_BT\lesssim \hbar/\tau$, are governed by singular diffusive particle-hole propagators\;\cite{AAbook}, $\mathcal{D}(q,\omega)= 1/(D_0q^2+\omega)$. (Here $D_0$ is the diffusion constant proportional to the elastic scattering rate $1/\tau$.) 
%.  
The in-plane magnetic field suppresses the singularity in the spin-triplet channels $(S_z=\pm 1)$ with opposite spin projections by introducing a gap proportional to the Zeeman energy scale $\Delta_z=g\mu_B B_\parallel$. This suppression gives rise to a negative magnetconductance, $\Delta\sigma(B_\parallel,T)=\sigma(B_\parallel,T)-\sigma(0,T) <0$ in the weak-field limit\;\cite{kawabata_zeeman,tvr}, $\Delta_z\lesssim k_BT$. In the high-field limit,  $\Delta_z \gg k_BT$,  the spin-bands are well split and the transport is governed entirely by the $S_z=0$ channels which are insensitive to spin-splitting\;\cite{sasha83b,sasha84b,pedagogical}.  In a multivalley system, valley splitting, $\Delta_v$, and intervalley scattering $\Delta_\perp\equiv\hbar/\tau_\perp$, introduces additional gaps\;\cite{fukuyama1,fukuyama2}   further suppressing $\sigma(B_\parallel,T)$. The interplay of spin and valley degrees of freedom in the presence of finite  $\Delta_v$ and  $\Delta_\perp$ are studied in this paper. 

It is now well understood that the singular nature of $\mathcal{D}(q,\omega)$ leads to a strong enhancement of the \textit{e-e} scattering amplitudes  at low energies\;\cite{sasha83}. In two dimensions, renormalization group (RG) theory applied to a weakly disordered system has been extremely successful at capturing this scale dependence  to all orders in the \textit{e-e} scattering amplitudes\;\cite{yellowbook,pedagogical}. 
Strong \textit{e-e} scattering and energy renormalization effects, where the latter takes into account the renormalization of the Stoner enhancement factor, were incorporated into $\Delta\sigma(B_\parallel,T)$ in Refs.\;\onlinecite{raimondi_zeeman} and \onlinecite{castellani98}.
They are generalized here to include the effects of  $\Delta_v$ and $\Delta_\perp$ in a two-valley system. [It is assumed  that $\Delta_\perp < \Delta_v$, which appears to be the case in high mobility silicon inversion layers\;\cite{ZNAfits_vitkalov,tau_perp_gershenson}.]
The relevant interaction amplitudes are identified and the corresponding scaling equations are determined in  low, $\Delta_z\lesssim \Delta_\perp$, and  intermediate-fields, $\Delta_\perp\lesssim \Delta_z\lesssim \Delta_v$. The high-field limit $\Delta_z > \Delta_v$,  has been studied in great detail in Ref.\;\onlinecite{burmistrov_spinvalley}. [Note that for $k_BT \lesssim \Delta_z, \Delta_v$, the results in the high and intermediate field regimes are the same provided the spin and valley indices are interchanged.]  
%
%Our results  in the $\Delta_\perp \lesssim \Delta_z\lesssim\Delta_v$ regime are in complete agreement with Ref.\;\onlinecite{burmistrov_spinvalley} after interchanging the spin and valley indices.  

 %[The notations used in this paper are detailed in Ref.\;\onlinecite{punnoose_valley_RG}.]

The RG equations for finite $\Delta_v$ and $\Delta_\perp$ with $\Delta_z=0$  are detailed  in Ref.\;\onlinecite{punnoose_valley_RG}.   For finite $\Delta_z$,  it is convenient to work in the (spin-singlet/triplets)$\otimes$(valley-singlet/triplets) basis  so that the spin and valley degrees of freedom  are treated on equal footing in the particle-hole channels.  The diffusion propagators and the \textit{e-e} scattering amplitudes in this basis are discussed below.

\textit{Diffusion modes}: For $\Delta_z=0$, it was sufficient to label the modes  in terms of the valley states  $\mathcal{D}_\alpha(q,\omega)$, where $\alpha=\pm$ and $\perp$. (See Ref.\;\onlinecite{punnoose_valley_RG} for further details.)  $\alpha=+$ refers to the valley-singlet channel which is gapless, and  $\alpha=-$ and $\perp$ refer to the gapped valley-triplet channels with gaps proportional to $\Delta_\perp$ and $\Delta_v+\Delta_\perp$, respectively. Since $\mathcal{D}_-$  involves scattering only within  the same valley, it  is  insensitive to the splitting $\Delta_v$. It, however,  develops a gap $\Delta_\perp$ when intervalley scattering mixes the two valleys.  All modes are effectively gapless at high temperatures, $k_BT\gtrsim \Delta_v, \Delta_\perp$. [Temperature units $T_v, T_\perp$ and $T_z$ will be used interchangeably in the following to represent the scales $\Delta_v, \Delta_\perp$ and $\Delta_z$, respectively.]

For $\Delta_z\neq 0$,  the spin-triplet channels $S_z=\pm 1$  develop spin-gaps proportional to $\Delta_z$.  Hence, it is convenient to label the propagators as $\mathcal{D}_{t\alpha}$ and $\mathcal{D}_{s\alpha}^\sigma$, where the subscript $t$ corresponds to the spin-triplet channels with $S_z=\pm 1$, and $s$ labels the $S_z=0$ channels, with the singlet and the triplet $S_z=0$ channels labeled by $\sigma=\pm$.

\textit{e-e scattering amplitudes}: In a single valley system, the \textit{e-e} scattering amplitudes are uniquely described  by the spin texture of the scattering channel. The amplitudes $\Gamma_s$ and $\Gamma_t$ are used to describe the scattering of particle-hole pairs in the spin-singlet and triplet channels, respectively.  They are related to the standard static Fermi-liquid amplitudes  $\Gamma_1$ and $\Gamma_2$ as  $\Gamma_s=\Gamma_1-\Gamma_2/2$ and $\Gamma_t=-\Gamma_2/2$.   These definitions are easily extended to\;\cite{punnoose_valley_RG},  $\Gamma_{s\alpha}^\sigma=\Gamma_{2\alpha}-4\Gamma_{1\alpha}^\sigma$  and $\Gamma_{t\alpha}=\Gamma_{2\alpha}$, where $\alpha=\pm,\perp,$ and $\sigma=\pm$. [For notational convenience, the amplitudes $\Gamma_{s\alpha}^\sigma$ are defined with a factor of -4.]  Note that the intervalley scattering amplitudes $\Gamma_{1\perp}^\sigma$  are generally negligibly small in a clean system because the Coulomb scattering involving large momentum $Q_0$ perpendicular to the 2D plane is suppressed when the width of the inversion layer is many times larger than the lattice  spacing, hence $\Gamma_{s\perp}^\sigma=\Gamma_{t\perp}=\Gamma_{2\perp}$.   Together, the total number of amplitudes equal $\Gamma_{s\alpha}^\sigma\{4\}+\Gamma_{t\alpha}\{4\}+\Gamma_{s\perp}^\sigma\{4\}+\Gamma_{t\perp}\{4\}=\{16\}$; the number in brackets denotes the number of channels.

\begin{figure}[h]
\includegraphics[width=0.95\linewidth]{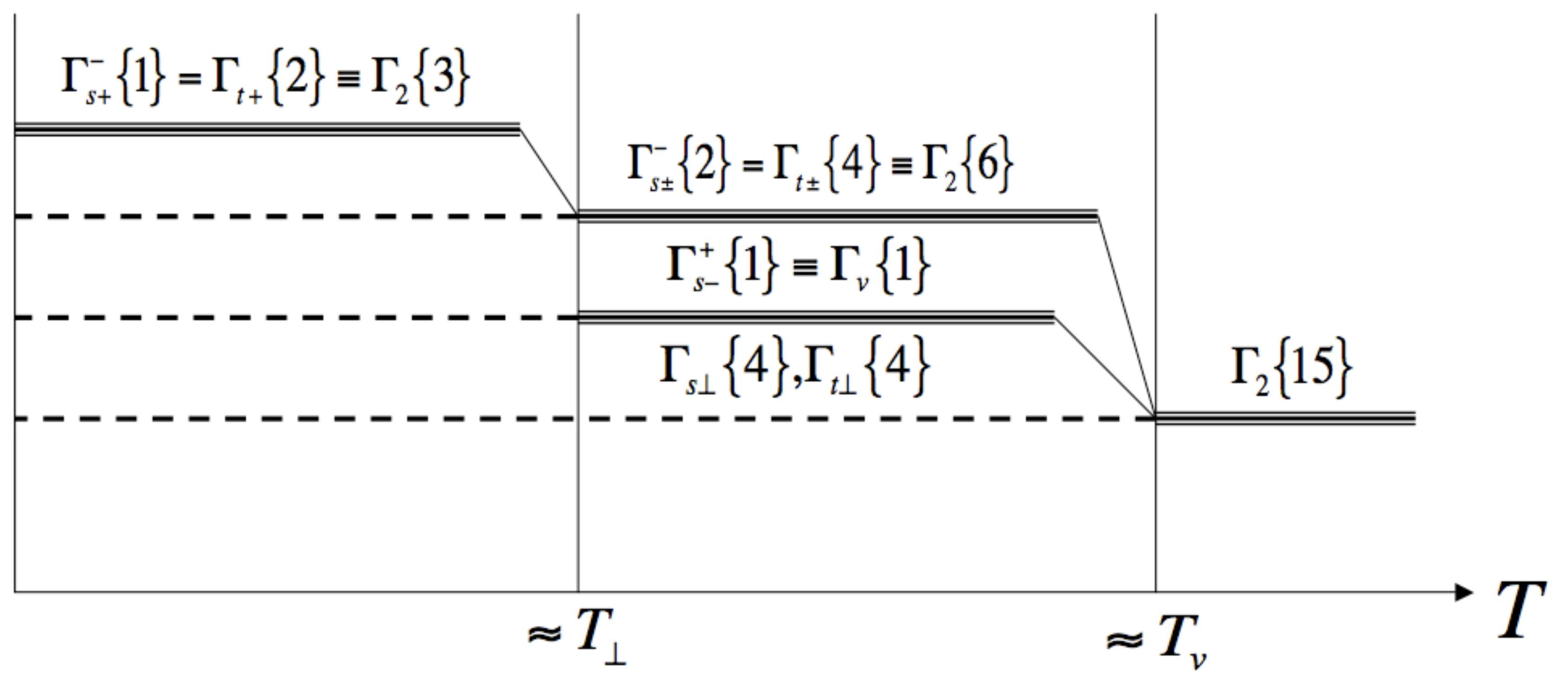}
 \caption{Schematics showing the classification of the \textit{e-e} scattering amplitudes as a function of temperature $T$  in the presence of valley splitting $T_v$ and intervalley scattering $T_\perp$. The relevant amplitudes are marked by solid lines with the degenerate amplitudes grouped together, while dashed lines mark the irrelevant amplitudes.   The spin and valley structure of the  amplitudes are described in the main text. }
 \label{fig:zero_field_gamma}
\end{figure}

In the high temperature limit, $T\gtrsim T_v,T_z$, the amplitudes $\Gamma_{1\alpha}^\sigma$, except for $\Gamma^+_{1+}$, are identically zero. The $\Gamma_{1+}^+$ amplitude, which involves scattering in the  spin and valley singlet channels, (spin-singlet)$\otimes$(valley-singlet), is special in that it combines with the long-ranged part of the Coulomb interaction to produce a universal amplitude\;\cite{AAbook}. [Details are given below in section \ref{sec:Tperp_Tz_T_Tv}.]  Hence, all 15  of the 16 amplitudes are equal and evolve as $\Gamma_2$. They  are shown grouped together when $T\gtrsim T_v$ in Fig.\;\ref{fig:zero_field_gamma}. 

When $T_\perp\lesssim T \lesssim T_v$, the $\mathcal{D}_\perp$ propagators are gapped, the corrections to $\Gamma_\perp$ are therefore non-singular and hence irrelevant. On the other hand, the  $\Gamma_{1-}^+$ amplitude in the (spin-singlet)$\otimes$(valley-triplet) channel, which  vanishes at high temperatures, was shown in Ref.\;\onlinecite{punnoose_valley_RG} to be generated under the RG transformations when $T\lesssim T_v$.  (To emphasize that $\Gamma_{s-}^+$  arises as an independent scaling variable only when the valley sub-bands are split, it is designated as $\Gamma_v\equiv \Gamma_{s-}^+$.) This is a generic feature of mutliband systems with sub-band splittings, it was first discussed in Ref.\;\onlinecite{burmistrov_spinvalley} in the opposite case  $T_v\lesssim T\lesssim T_z$ where the  relevant amplitude is $\Gamma_{1+}^-$, with the spin and valley indices interchanged.

The splitting of the 15 amplitudes below $T_v$  are  shown schematically in Fig.\;\ref{fig:zero_field_gamma}. The solid horizontal lines  mark the relevant amplitudes while the dashed lines mark the irrelevant ones. The  degenerate amplitudes under the RG flow are grouped together with the degeneracy indicated in curly brackets.
At the lowest temperature $T\lesssim T_\perp$,  when the two valleys are strongly mixed,  only the valley-singlet propagator $\mathcal{D}_+$ remains gapless. Hence, only the amplitudes in the valley-singlet channel $\Gamma_+$  survives.

Clearly, the number of relevant \textit{e-e} scattering amplitudes in a multiband system at a given scale is sensitive to the splitting and the interband scattering rates within the bands. Calculations in different temperature regimes in the presence of spin-splitting are detailed below. 
[Since it is seen experimentally that the phase breaking rate saturates at low temperatures for low electron densities\;\cite{dephasing_kravchenko}, where the results obtained in this paper are most relevant, the contribution from the cooperon (particle-particle) channel has been suppressed in the calculations.]
 %
  
%\section{Scaling equations}

%\subsection{Intermediate-field range: $T_\perp\lesssim T_z\lesssim T_v$}

\begin{figure}[h]
\includegraphics[width=0.95\linewidth]{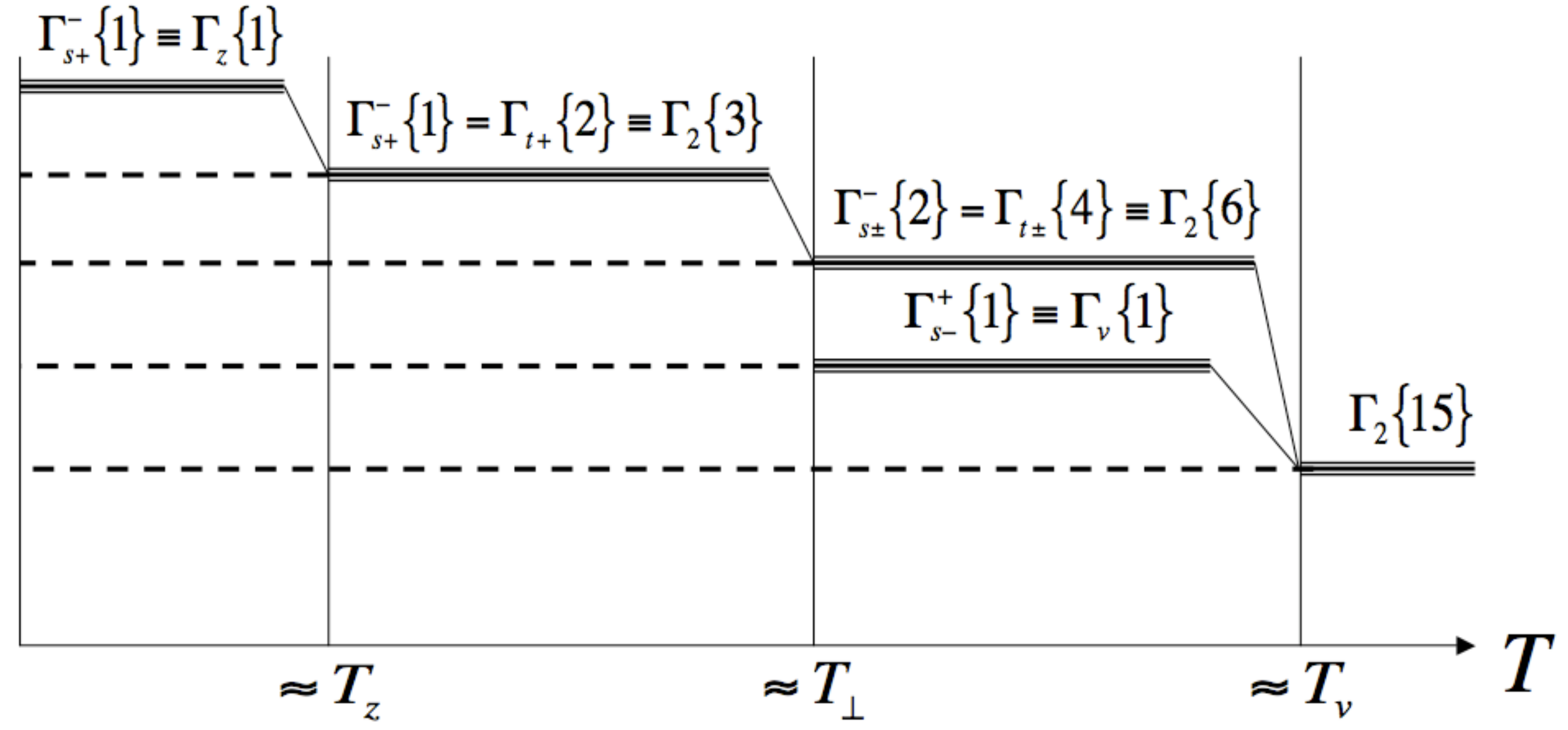}
 \caption{Schematics showing the classification of the relevant \textit{e-e} scattering amplitudes for weak spin-splitting, $T_z\lesssim T_\perp,T_v$, as function of temperature.   } \label{fig:Tz-Tperp}
\end{figure}

Fig.\;\ref{fig:Tz-Tperp} shows schematically the effect of  a weak magnetic field $T_z\lesssim T_\perp, T_v$ on the amplitudes. The spin gap suppresses the singular corrections in the spin-triplet channels, hence only $\Gamma_{s+}^-$  (and $\Gamma_{s+}^+$) develops singular diffusion corrections. The amplitude is designated as $\Gamma_z\equiv\Gamma_{s+}^-$ to emphasize that $T\lesssim T_z$. 

\begin{figure}[h]
\includegraphics[width=0.95\linewidth]{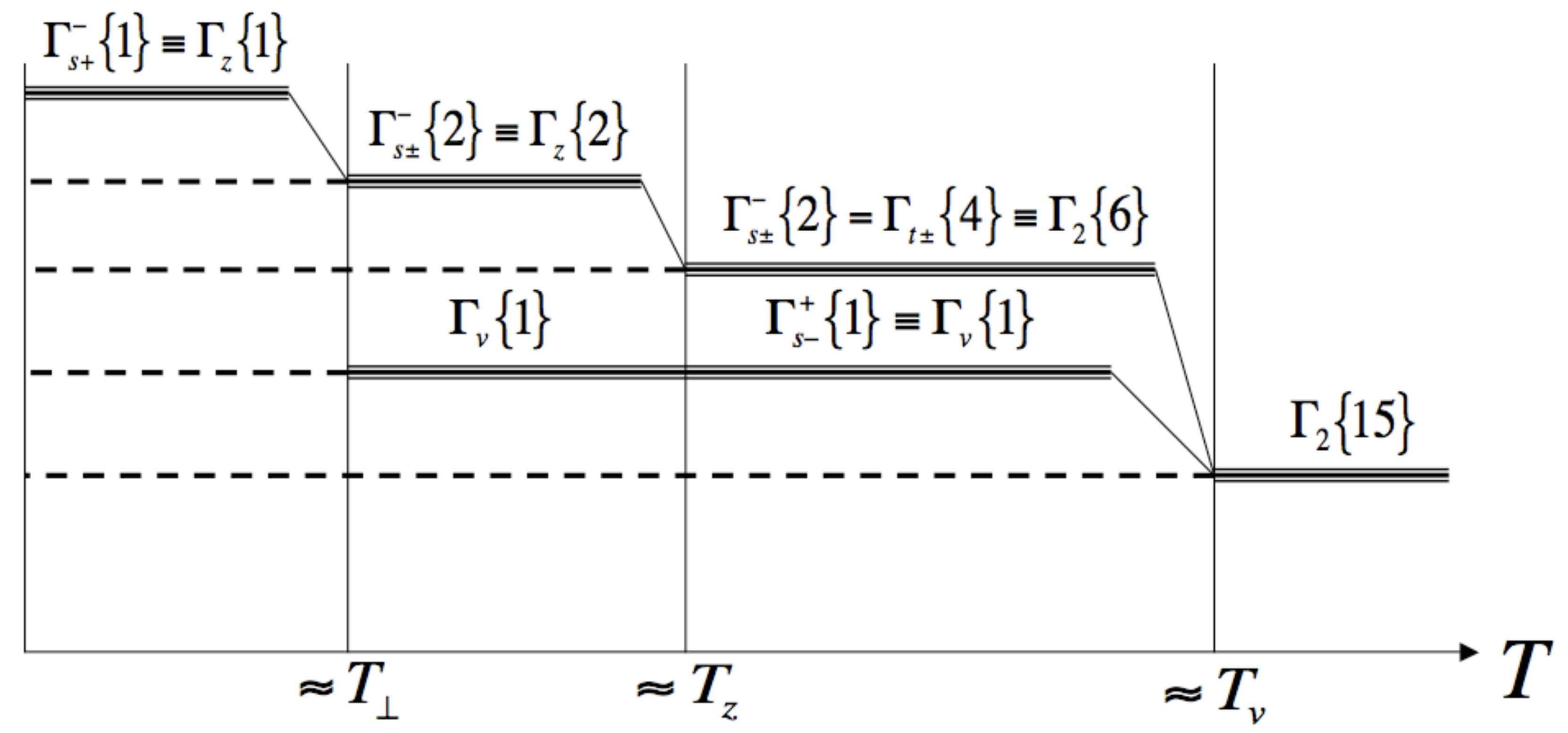}
 \caption{Schematics showing the classification of the relevant \textit{e-e} scattering amplitudes for intermediate values of the spin-splitting, $T_\perp\lesssim T_z\lesssim T_v$, as function of temperature.  } \label{fig:Tperp-Tz-Tv}
\end{figure}

Fig.\;\ref{fig:Tperp-Tz-Tv} shows  schematically   the relevant amplitudes for intermediate values of the spin-splitting  $T_\perp\lesssim T_z\lesssim T_v$. As in Fig.\;\ref{fig:Tz-Tperp}, the  spin-triplet channels, $\Gamma_t$, are irrelevant below $T_z$ due to the gap in the $\mathcal{D}_t$ propagators. As a result, the number of relevant amplitudes (not all degenerate) reduces from seven for $T\gtrsim T_z$ to three for $ T\lesssim T_z$.  

\section{Scaling equations}\label{scalingequations}

The RG equations in each of the temperature intervals shown in Figs.\;\ref{fig:Tz-Tperp} and \ref{fig:Tperp-Tz-Tv} are derived below. The relevant equations  when spin-splitting can be ignored, $T\gtrsim T_z$, have been derived in detail in Ref.\;\onlinecite{punnoose_valley_RG}. The logarithmic corrections are presented here in sections \ref{sec:Tperp_Tz_T_Tv} and \ref{sec:Tz_T_Tperp_Tv} after including the spin degrees of freedom explicitly. 

\subsection{$T_\perp,T_z\lesssim T\lesssim T_v$}\label{sec:Tperp_Tz_T_Tv}
Since the $\mathcal{D}_\perp$ modes are gapped for $T\lesssim T_v$,  their contributions are non-singular and hence dropped. All other modes are effectively gapless when $T\gtrsim T_z, T_\perp$. The gapless propagators are set equal to  $\mathcal{D}^\sigma_{s\pm}=\mathcal{D}_{t\pm}\equiv\mathcal{D}(q,\omega)=1/(Dq^2+z\omega)$, where $D$ is the renormalized diffusion constant and  $z$ parametrizes  the relative scaling of the frequency with respect to the length scale\;\cite{sasha83,sasha84}.  Both $D$ and $z$ acquire diffusion corrections in an interacting system. ($z=1$ for a non-interacting system\;\cite{AAbook}.) 

The nature  of the gapless diffusion modes induce the following relations on the amplitudes:   $\Gamma_{t+}=\Gamma_{t-}\equiv \Gamma_2$,  $\delta\Gamma_{1+}^+=\delta\Gamma_{1-}^+$ and $\delta\Gamma_{1\pm}^-=0$. Since  $\delta\Gamma_{1\pm}^-=0$, the amplitudes  $\Gamma_{s\pm}^-=\Gamma_2$ are degenerate. The  diffusion corrections in terms of these variables take the form\;\cite{punnoose_valley_RG}

\begin{subequations}\label{eqn:corrections_medT}
\begin{eqnarray}
%D
\frac{\delta D}{D}&=&-\frac{4}{\nu}\iint \left(\Gamma_{1-}^{+}+\Gamma_{1+}^{+}-2\Gamma_2\right)\mathcal{D}^3(q,\omega)Dq^2,\label{eqn:medT_D}\\
%z
\delta z&=&-\frac{1}{\pi\nu}\int\frac{d^2q}{(2\pi)^2} \left(\Gamma_{1-}^{+}+\Gamma_{1+}^{+}-2\Gamma_2\right)\mathcal{D}(q,0),\label{eqn:medT_z}\\
%gamma_2
\delta\Gamma_2&=&\frac{1}{\pi\nu}\int\frac{d^2q}{(2\pi)^2}\left(\Gamma_{1-}^{+}+\Gamma_{1+}^{+}\right)\mathcal{D}(q,0)+{8}\Psi(\Gamma_2),\hspace{0.65cm}\label{eqn:medT_g2}\\
%gamma_1
\delta\Gamma_{1\pm}^{+}&=&\frac{1}{2\pi\nu}\int\frac{d^2q}{(2\pi)^2}\Gamma_2\mathcal{D}(q,0)+{2}\Psi(\Gamma_2).\label{eqn:medT_g1}
\end{eqnarray}
\end{subequations}
The single integral is defined as $\int=d^2q/(2\pi)^2$ and the double integral as $\iint=\int d^2q/(2\pi)^2\int d\omega/(2\pi)$. The density of states  per spin and valley $\nu=m/2\pi$. %
The  contributions of the ``ring" diagrams\;\cite{sasha83}  equals  $\Psi(\Gamma_2)$, where (see Fig.\;5 in \onlinecite{punnoose_valley_RG})

\begin{eqnarray}\label{eqn:Psi}
\Psi(\Gamma_2)&=&
+\frac{1}{\nu}\iint\Gamma_2\left[\Gamma_2\mathcal{D}^2\right]-\frac{1}{2}\left[\Gamma_2^2\mathcal{D}^2\right]\nonumber\\
&&-\frac{1}{\nu}\iint\omega\Gamma_2
\left[\Gamma_2^2\mathcal{D}^3\right]-\omega\Gamma_2^2\left[\Gamma_2\mathcal{D}^3\right]\nonumber\\
&&-\frac{1}{2\nu}\iint\omega^2\Gamma_2^2\left[\Gamma_2^2\mathcal{D}^4\right].
\end{eqnarray}

As noted already, the relevance of the $\Gamma_{1-}^+$ amplitude in the temperature range  $ T_\perp\alt T\alt T_v$ is specific to problems with split-bands in a multivalley system.  Although the corrections $\delta \Gamma_{1+}^{+}=\delta\Gamma_{1-}^{+}$  for $T\lesssim T_v$, their initial values are different. The amplitude  $\Gamma_{1-}^+=0$ when $T\gtrsim T_v$, while the singlet amplitude $\Gamma_{1+}^+$ is special as it combines with the static limit of the Coulomb interaction, denoted here as $\Gamma_{0+}^+$  (it is conventionally denoted simply as $\Gamma_0$ in a single valley system with degenerate spin bands\;\cite{AAbook}). The $\Gamma_{1+}^+$ amplitude appearing in (\ref{eqn:medT_D})-(\ref{eqn:medT_g2}) are to be replaced by its long-ranged value
\begin{equation}\label{eqn:gamma1_LR}
\Gamma_{1+}^+\longrightarrow\Gamma^{LR}_1=\Gamma_{0+}^++\Gamma_{1+}^+.
\end{equation}
When combined with the $\Gamma_{2}$ amplitude, the long-ranged singlet amplitude is given as: $\Gamma_{s}^{LR}=\Gamma_2-4\Gamma_1^{LR}$. (To be consistent with the notations in this paper, $\Gamma_{s}^{LR}$ is defined with an extra factor of -4.)
It is easily verified by combining Eqs.\;(\ref{eqn:medT_z})-(\ref{eqn:medT_g1})  that  the singlet combination  $\delta (z +\Gamma_{s}^{LR})=0$ is satisfied at all length scales, provided the corrections to the static amplitude $\delta \Gamma_{0+}^+=0$.  This is  a well established result, with great importance for the general structure of the theory\;\cite{sasha83,yellowbook}.

Having obtained the leading logarithmic corrections, the scaling equations are derived to first order in the dimensionless resistance $\rho=1/4(2\pi^2\nu D)$ and to all orders in the \textit{e-e} interaction amplitudes  by performing the ladder summations described in Fig.\;6 in Ref.\;\onlinecite{punnoose_valley_RG}.  It amounts to replacing the static  amplitudes $\Gamma_i$ by the dynamical amplitudes $U_i(q,\omega)$:

\begin{equation}\label{eqn:U}
U_{i}(q,\omega)=\Gamma_{i} \frac{\mathcal{D}_{i}(q,\omega)}{\mathcal{D}(q,\omega)},
\end{equation}
where, the propagators $\mathcal{D}_i$ are defined as

\begin{equation}
\mathcal{D}_{i}(q,\omega)=\frac{1}{Dq^2+(z+\Gamma_{i})\omega}.
\end{equation}

The amplitudes $\Gamma_i$ represents $\Gamma_2, \Gamma_{s-}^+$ and $\Gamma_{s}^{LR}$. Note that since  the leading logarithmic corrections involve only one momentum integration, it generates only one factor of $1/D$. The corrections are therefore limited to the first order in resistance $\rho$ (disorder). The limitation on the number of momentum integrations constraints the number of \textit{e-e} vertices in the skeleton diagrams. The ladder sums extend the skeleton diagrams to all orders in $\Gamma_i$ without changing the number of momentum integrations. Note, however, that only those interaction vertices involving frequency integrations can be extended to include dynamical effects. 
These amplitudes are enclosed in square brackets in Eq.\;(\ref{eqn:Psi}).  Substituting the $\Gamma_2$ amplitudes in the square brackets with $U_2$ and performing the $q,\omega$ integrals leads to the very simple expression\;\cite{sasha83,pedagogical}:

\begin{equation}\label{eqn:psi_integral}
\Psi(\Gamma_2)=\left(\frac{\Gamma_2^2}{z}\right)\times \frac{\rho}{2}\log\left(\frac{1}{T\tau}\right).
\end{equation} 

The remaining single integrals  $\int d^2q \mathcal{D}(q,0)$ involving only momentum integrations are easily evaluated to give

\begin{equation}\label{eqn:rhointegral}
\frac{1}{\pi\nu}\int \frac{d^2q}{(2\pi)^2}\mathcal{D}(q,0)=2\rho\log\left(\frac{1}{T\tau}\right).
\end{equation}

The integrals in $\delta D$ containing $\omega$ integrations  remain to be evaluated.
Before the integrals  can be done, the  $\Gamma_{1+}^+$ amplitude is  replaced with $\Gamma_{1}^{LR}$ following Eq.\;(\ref{eqn:gamma1_LR}), after which the amplitudes $\Gamma_{1}^{LR}, \Gamma_{1-}^+$ and $\Gamma_2$  are rearranged to form $\Gamma_{s}^{LR}$ and $\Gamma_{s-}^+$ and $\Gamma_2$ and then extended to $U_s^{LR}, U_{s-}^+$ and $U_2$, respectively.

When the equation for $\rho$ is expressed in terms of the scaling variables,  $\gamma_2=\Gamma_2/z$ and $\gamma_v=\Gamma_{s-}^+/z$, the equations for  $\rho$, $\gamma_2$ and $\gamma_v$ form a closed set of  equations  independent of $z$. 
The final RG equations in the range $T_\perp,T_z\lesssim T\lesssim T_v$ 
are given below with the scale $\xi$ defined to logarithmic accuracy as $\xi=\log(1/T\tau)$.
\begin{subequations}\label{eqn:RG_medT}
\begin{eqnarray}
\frac{d\rho}{d\xi}&=& \rho^2\left(1-\Phi(\gamma_v)-6\Phi(\gamma_2)\right),\\
\frac{d\gamma_2}{d\xi}&=&\frac{\rho}{2}\left[(1+\gamma_2)^2+(1+\gamma_2)(\gamma_2-\gamma_v)\right],\\
\frac{d\gamma_v}{d\xi}&=&\frac{\rho}{2}(1+\gamma_v)(1-\gamma_v-6\gamma_2),\\
\frac{d\ln z}{d\xi}&=&-\frac{\rho}{2}\left(1-\gamma_v-6\gamma_2\right).
\end{eqnarray}
\end{subequations}
The function $\Phi(\gamma)$ is defined as
\begin{equation}\label{eqn:Phi}
\Phi(\gamma)=\left(1+\frac{1}{\gamma}\right)\log(1+\gamma)-1.
\end{equation} 

As described in Fig.\;\ref{fig:Tperp-Tz-Tv}, the 15 degenerate amplitudes for $T\gtrsim T_v$ split into six $\Gamma_2$ and one $\Gamma_v$ amplitude when $T\lesssim T_v$. [This splitting of the amplitudes is generic to multiband systems with subband  splittings. The same equations are obtained when instead of the valley bands, the spin bands are split\;\cite{burmistrov_spinvalley}, i.e., $T_v\lesssim T\lesssim T_z$.] Note that $\gamma_v$ coincides with $\gamma_2$ when $T\approx T_v$.

\subsection{$T_z\lesssim T\lesssim T_\perp,T_v$}\label{sec:Tz_T_Tperp_Tv}

 The relevant amplitudes in the presence of strong valley mixing ($T\lesssim T_\perp$) correspond to scattering in the valley-singlet channels, $\Gamma_{s+}^-$ and $\Gamma_{t+}$ . Since $\delta\Gamma_{1+}^-=0$ vanishes in the absence of spin-splitting ($T\gtrsim T_z$), it follows that the amplitudes $\Gamma_{s+}^-=\Gamma_{t+}=\Gamma_2$ are all equal and satisfy the equations\;\cite{punnoose_valley_RG}
\begin{subequations}\label{eqn:corrections_lowT}
\begin{eqnarray}
%D
\frac{\delta D}{D}&=&-\frac{4}{\nu}\iint \left(\Gamma_{1+}^+-\Gamma_2\right)\mathcal{D}^3(q,\omega)Dq^2,
\label{eqn:corrections_lowT_D}\\
%z
\delta z&=&-\frac{1}{\pi\nu}\int\frac{d^2q}{(2\pi)^2} \left(\Gamma_{1+}^+-\Gamma_2\right)\mathcal{D}(q,0),
\label{eqn:corrections_lowT_z}\\
%gamma_2
\delta\Gamma_2&=&\frac{1}{\pi\nu}\int\frac{d^2q}{(2\pi)^2}\Gamma_{1+}^+\mathcal{D}(q,0)+{4}\Psi(\Gamma_2),
\label{eqn:corrections_lowT_g2}\\
%gamma_1
\delta\Gamma_{1+}^+&=&\frac{1}{4\pi\nu}\int\frac{d^2q}{(2\pi)^2}\Gamma_2\mathcal{D}(q,0)+\Psi(\Gamma_2).\label{eqn:corrections_lowT_g1}
\end{eqnarray}
\end{subequations}
The coefficient  of $\Gamma_2$   and the ring diagrams $\Psi(\Gamma_2)$ in (\ref{eqn:corrections_lowT_D})-(\ref{eqn:corrections_lowT_g2}) are suppressed by a factor two  when compared with (\ref{eqn:medT_D})-(\ref{eqn:medT_g2}) since they no longer contain a valley sum. The corrections to $\Gamma_{1\alpha}^\sigma$ in (\ref{eqn:corrections_lowT_g1}) already do not carry a valley sum, only  half the amplitude  involving the same valley, however, acquires  corrections when the valley-bands are mixed, which accounts for the overall factor of half when compared with (\ref{eqn:medT_g1}). 
Note that the condition $\delta(z+\Gamma_s^{LR})=0$ is satisfied. Following the procedure described in  section \ref{sec:Tperp_Tz_T_Tv}, the RG equations read

\begin{subequations}\label{eqn:RG_lowT}
\begin{eqnarray}
\frac{d\rho}{d\xi}&=& \rho^2\left(1-3\Phi(\gamma_2)\right)\\
\frac{d\gamma_2}{d\xi}&=&\frac{\rho}{2}(1+\gamma_2)^2\\
\frac{d\ln z}{d\xi}&=&-\frac{\rho}{2}\left(1-3\gamma_2\right)
\end{eqnarray}
\end{subequations}

The function $\Phi(\gamma)$ is defined in Eq.\;(\ref{eqn:Phi}).
As described in Fig.\;\ref{fig:Tz-Tperp}, only three of the 15 degenerate amplitudes survive when $T\lesssim T_\perp$ when spin-splitting can be neglected  $T\gtrsim T_z$.  The high field cases are discussed below, i.e., $T\lesssim T_z$.

\subsection{$T_\perp\lesssim T\lesssim T_z\lesssim T_v$}\label{sec:Tperp_T_Tz_Tv}

It should be noted that the results for $T\lesssim T_z\lesssim T_v$  is equivalent to the situation if the gap scales were reversed, i.e., $T\lesssim T_v\lesssim T_z$, provided of course the spin and valley indices are interchangeable, which is the case when  $T\gtrsim T_\perp$. The RG equations for $T_v\lesssim T_z$ are derived in Ref.\;\onlinecite{burmistrov_spinvalley}. The opposite situation $T_z\lesssim T_v$ is derived here. 

When $T\lesssim T_z$, the $\mathcal{D}_{t\pm}$ propagators are gapped, and hence the corrections in the  $S_z=\pm 1$ channel are non-singular. The corresponding amplitudes $\Gamma_{t\pm}$ are therefore irrelevant at these temperatures, which reduces the number of relevant interaction amplitudes by four. Furthermore, the amplitude $\Gamma_{1+}^-$ acquires diffusion corrections\;\cite{burmistrov_spinvalley} when $T\lesssim T_z$ in the same way that $\Gamma_{1-}^+$ does when $T\lesssim T_v$.  Since $T\lesssim T_z$ and $T_v$, the amplitude $\Gamma_{1-}^-$ also acquires logarithmic corrections.  As a result, both  $\Gamma_{s\pm}^-$ are different from $\Gamma_2$ when $T\lesssim T_z,T_v$. 
After including the contributions from $\Gamma_{1\alpha}^\sigma$, the diffusion corrections for $T\lesssim T_z$   take the form:
\begin{subequations}\label{eqn:corrections_Tperp_T}
\begin{eqnarray}
%D
\label{corrections_Tperp_T_D}
\frac{\delta D}{D}&=&-\frac{4}{\nu}\iint \left(\sum_{\alpha,\sigma=\pm} \Gamma_{1\alpha}^\sigma-\Gamma_2\right)\; \mathcal{D}^3(q,\omega)Dq^2,\hspace{0.75cm}\\
%z
\label{corrections_Tperp_T_z}
\delta z&=&-\frac{1}{\pi\nu}\int \left(\sum_{\alpha,\sigma=\pm} \Gamma_{1\alpha}^\sigma-\Gamma_2\right)\; \mathcal{D}(q,0),\\
%gamma_2
\label{corrections_Tperp_T_g2}
\delta\Gamma_2&=&\frac{1}{\pi\nu}\int \sum_{\alpha,\sigma=\pm} \Gamma_{1\alpha}^\sigma \; \mathcal{D}(q,0)+{4}\Psi(\Gamma_2),\\
%gamma_1
\label{corrections_Tperp_T_g1}
\delta\Gamma_{1\alpha}^\sigma&=&\frac{1}{4\pi\nu}\int\Gamma_2\mathcal{D}(q,0)+\Psi(\Gamma_2).\end{eqnarray}
\end{subequations}

The coefficient  of $\Gamma_2$   and the ring diagrams $\Psi(\Gamma_2)$ in (\ref{corrections_Tperp_T_D})-(\ref{corrections_Tperp_T_g2}) are suppressed by a factor two  when compared with (\ref{eqn:medT_D})-(\ref{eqn:medT_g2}) since they no longer contain a spin sum. The corrections to $\Gamma_{1\alpha}^\sigma$ in (\ref{corrections_Tperp_T_g1}) already do not carry a spin sum. Only  half the amplitude  involving the same spin, however, acquires  corrections when the spin-bands are split, which accounts for the overall factor of half when compared with (\ref{eqn:medT_g1}).  Since  $\delta\Gamma_{1+}^-=\delta\Gamma_{1-}^-$,  the amplitudes, after combining with $\Gamma_2$, can be grouped together as  $\Gamma_{s\alpha}^-\equiv\Gamma_z$. Extending the singlet amplitude $\Gamma_{1+}^+$ to include the static long ranged part of the Coulomb interactions $\Gamma_{1}^{LR}$ as discussed in Eq.\;(\ref{eqn:gamma1_LR}) and using the identity $\Gamma_2-\sum_{\alpha,\sigma}\Gamma_{1\alpha}^\sigma=\sum_{\alpha,\sigma}\Gamma_{s\alpha}^\sigma/4$,  the amplitude $\Gamma_2$ can be eliminated from  Eqs.\;(\ref{corrections_Tperp_T_D})-(\ref{corrections_Tperp_T_g1}) in favor of the amplitudes $\Gamma_{s}^{LR}, \Gamma_z$ and $\Gamma_v$  as

\begin{subequations}\label{eqn:Tperp_T}
\begin{eqnarray}
%D
\label{Tperp_T_D}
\frac{\delta D}{D}&=&\frac{1}{\nu}\iint (\Gamma_s^{LR}+\Gamma_v+2\Gamma_z) \mathcal{D}^3(q,\omega)Dq^2,\\
%z
\label{Tperp_T_z}
\delta z&=&\frac{1}{4\pi\nu}\int (\Gamma_s^{LR}+\Gamma_v+2\Gamma_z)\mathcal{D}(q,0),\\
%gamma_
\label{Tperp_T_g}
\delta\Gamma_z&=&\delta\Gamma_v=\delta\Gamma_s^{LR}=-\delta z.\end{eqnarray}
\end{subequations}

Combining (\ref{corrections_Tperp_T_g2}) and (\ref{corrections_Tperp_T_g1}) to give Eq.\;(\ref{Tperp_T_g}) is possible only because the $\Psi(\Gamma_2)$ contribution cancels exactly when the  sum over opposite spin-projections  are  suppressed due to spin-splitting\;\cite{sasha84b}. Also note in (\ref{Tperp_T_g}), that the singlet combination $\delta(z+\Gamma_s^{LR})=0$ holds explicitly, as needed for the consistency of the RG theory\cite{yellowbook,pedagogical}. 

The RG equations are obtained by evaluating the integrals after extending the static amplitudes by the dynamical amplitudes $U_i$ defined in Eq.\;(\ref{eqn:U}). The RG equations  for $T_\perp\lesssim T\lesssim T_z\lesssim T_v$ are
\begin{subequations}\label{eqn:RG_Tperp_T}
\begin{eqnarray}
\frac{d\rho}{d\xi}&=& \rho^2\left(1-\Phi(\gamma_v)-2\Phi(\gamma_z)\right),\label{eqn:RG_Tperp_T_D}\\
\frac{d\gamma_z}{d\xi}&=&\frac{\rho}{2}(1+\gamma_z)\left(1-\gamma_v-2\gamma_z\right),\label{eqn:RG_Tperp_T_z}\\
\frac{d\gamma_v}{d\xi}&=&\frac{\rho}{2}(1+\gamma_v)(1-\gamma_v-2\gamma_z),\label{eqn:RG_Tperp_gv}\\
\frac{d\ln z}{d\xi}&=&-\frac{\rho}{2}\left(1-\gamma_v-2\gamma_z\right).\label{eqn:RG_Tperp_T_z}
\end{eqnarray}
\end{subequations}

As described in Fig.\;\ref{fig:Tperp-Tz-Tv}, the four $\Gamma_{t\pm}$ amplitudes are suppressed when $T\lesssim T_z$, leaving two  $\Gamma_z$ amplitudes, which  evolve away from $\Gamma_2$.   Note that $\gamma_z\approx \gamma_2$ when $T\approx T_z$, while $\gamma_v\approx\gamma_2$ when $T\approx T_v$. [The RG equations when spin-splitting is large, $T\lesssim T_v\lesssim T_z$, take the same form as Eqs.\;(\ref{eqn:RG_Tperp_T_D})-(\ref{eqn:RG_Tperp_T_z}) provided the spin and valley indices are interchanged; see Ref.\;\onlinecite{burmistrov_spinvalley} for details.]
   
\subsection{$T\lesssim T_\perp,T_z,T_v$}\label{sec:T_Tperp_Tz_Tv}

The two valleys are strongly mixed when $T\lesssim T_\perp$,  leaving only the valley-singlet propagators $\mathcal{D}_{s+}^\sigma$  gapless. Hence, only  $\Gamma_{s+}^-=\Gamma_z$  and  $\Gamma_{1+}^+$, survive at low temperatures. The corresponding diffusion corrections read

\begin{subequations}\label{eqn:corrections_T_Tperp}
\begin{eqnarray}
%D
\label{corrections_T_Tperp_D}
\frac{\delta D}{D}\!\!&=&\!\!-\frac{4}{\nu}\iint \left(\Gamma_{1+}^-+\Gamma_{1+}^+-\frac{1}{2}\Gamma_2\right)\; \mathcal{D}^3(q,\omega)Dq^2,\hspace{0.85cm}\\
%z
\label{corrections_T_Tperp_z}
\delta z&=&-\frac{1}{\pi\nu}\int  \left(\Gamma_{1+}^-+\Gamma_{1+}^+-\frac{1}{2}\Gamma_2\right)\; \mathcal{D}(q,0),\\
%gamma_2
\label{corrections_T_Tperp_g2}
\delta\Gamma_2&=&\frac{1}{\pi\nu}\int  \left(\Gamma_{1+}^-+\Gamma_{1+}^+\right)\mathcal{D}(q,0)+{2}\Psi(\Gamma_2),\\
%gamma_1
\label{corrections_T_Tperp_g1}
\delta\Gamma_{1+}^\sigma&=&\frac{1}{8\pi\nu}\int\Gamma_2\mathcal{D}(q,0)+\frac{1}{2}\Psi(\Gamma_2)\end{eqnarray}
\end{subequations}

All terms involving $\Gamma_2$ amplitudes are suppressed by a factor of two in (\ref{corrections_T_Tperp_D})-(\ref{corrections_T_Tperp_g2}) compared to (\ref{corrections_Tperp_T_D})-(\ref{corrections_Tperp_T_g2}) due to the suppression of the $\Gamma_{t-}$ amplitudes, which are irrelevant when $T\lesssim T_\perp$. The equations can be  simplified in terms of the amplitudes $\Gamma_{s}^{LR}$ and $\Gamma_{z}$ as
\begin{subequations}\label{eqn:T_Tperp}
\begin{eqnarray}
%D
\label{T_Tperp_D}
\frac{\delta D}{D}&=&\frac{1}{\nu}\iint (\Gamma_s^{LR}+\Gamma_z) \mathcal{D}^3(q,\omega)Dq^2,\hspace{0.5cm}\\
%z
\label{T_Tperp_z}
\delta z&=&\frac{1}{4\pi\nu}\int (\Gamma_s^{LR}+\Gamma_z)\mathcal{D}(q,0),\\
%gamma_
\label{T_Tperp_g}
\delta\Gamma_z&=&\delta\Gamma_s^{LR}=-\delta z.\end{eqnarray}
\end{subequations}

Note again that the condition $\delta(z+\Gamma_s^{LR})=0$ is satisfied. Following the procedure followed in the previous sections, the RG equations for $T\lesssim T_\perp,T_z,T_v$ are
\begin{subequations}\label{eqn:RG_T_Tperp}
\begin{eqnarray}
\frac{d\rho}{d\xi}&=& \rho^2\left(1-\Phi(\gamma_z)\right),\\
\frac{d\gamma_z}{d\xi}&=&\frac{\rho}{2}(1+\gamma_z)\left(1-\gamma_z\right),\\
\frac{d\ln z}{d\xi}&=&-\frac{\rho}{2}\left(1-\gamma_z\right).
\end{eqnarray}
\end{subequations}

These equations coincide with the results obtained in the case of a single valley with spin-splitting studied in Ref.\;\onlinecite{sasha84}. Strong intervalley scattering for $T\lesssim T_\perp$ mixes the two valleys to effectively produce a single valley. 

\section{Conclusions}
%\subsection{Low-field range: $T_z\lesssim T_\perp\lesssim T_v$}\label{sec:lowfield}

The derivation of the scaling equations  in Sec.\;\ref{scalingequations} were carried out keeping only the gapless valley and spin channels in each temperature interval. The scale dependence of the dimensionless resistance $\rho=(e^2/\pi h)R_\square$, where $R_\square$ is the sheet resistance, is then obtained by integrating the  self consistent set of  scaling equations separately in each temperature interval and matching the values of the amplitudes and resistance at the boundaries of each interval. Since the intervals are sensitive to the value of $T_z$, one obtains in this way $\rho(B_\parallel,T)$ whose inverse gives $\sigma(B_\parallel,T)=1/\rho(B_\parallel,T)$. This method is, however, not accurate as the crossover regions have finite contributions from the gapped channels and are hence non-universal. 

%[A method to obtain crossover RG equations has been recently suggested in Ref.\;\onlinecite{burmi_CRG}.] 

The case of weak spin-splitting $T_z\lesssim T$ can be treated fairly accurately, however. In this case the sensitivity to $B_\parallel$ arises only from the presence of a weak spin-gap in the triplet channels below the scale set by $T$. Hence, subtracting $\sigma(0,T)$ from $\sigma(B_\parallel,T)$ captures only the contributions originating from the suppression of the triplet channels.  The explicit form  $\Delta\sigma(B_\parallel,T)$ for the single valley case was derived in Refs.~\onlinecite{tvr,castellani98} in the limit $T_z\ll T$. It is straightforward to extend the results to include the valley degrees of freedom by keeping track of the number of $\Gamma_t$ amplitudes in a given temperature interval as\;\cite{punnoose07}
\begin{equation}
\Delta \sigma (B_\parallel,T)=-0.091\frac{e^{2}}{2\pi h}N_t\gamma_{2}\left(\gamma_{2}+1\right)(T_z/T)^{2}.
\label{eqn:linear}
\end{equation}
The number of relevant $\Gamma_t$ amplitudes, $N_t$ can be read off directly from Figs.\;\ref{fig:zero_field_gamma}, \ref{fig:Tz-Tperp} and \ref{fig:Tperp-Tz-Tv} as $N_t=2$ for $T_z\lesssim T\lesssim T_\perp$, $N_t=4$ for $T_z\lesssim T\lesssim T_v$ and $N_t=8$ for $T\gtrsim T_v, T_z$.    

To summarize, RG equations in the presence of spin-splitting induced by a parallel magnetic field have been obtained in a two valley system in the valley-split and strong intervalley scattering regimes. The behavior of the magnetoconductance in the limit of weak magnetic field are discussed.

\section{Acknowledgements}
The author would like to thank A. M. FinkelÕstein and S. V. Kravchenko for fruitful discussions. This work was supported  by DOE Grant No. DOE-FG02-84-ER45153.

%\bibliography{MITbib}

\begin{thebibliography}{20}
\expandafter\ifx\csname natexlab\endcsname\relax\def\natexlab#1{#1}\fi
\expandafter\ifx\csname bibnamefont\endcsname\relax
  \def\bibnamefont#1{#1}\fi
\expandafter\ifx\csname bibfnamefont\endcsname\relax
  \def\bibfnamefont#1{#1}\fi
\expandafter\ifx\csname citenamefont\endcsname\relax
  \def\citenamefont#1{#1}\fi
\expandafter\ifx\csname url\endcsname\relax
  \def\url#1{\texttt{#1}}\fi
\expandafter\ifx\csname urlprefix\endcsname\relax\def\urlprefix{URL }\fi
\providecommand{\bibinfo}[2]{#2}
\providecommand{\eprint}[2][]{\url{#2}}

\bibitem[{\citenamefont{Anissimova et~al.}(2007)\citenamefont{Anissimova,
  Kravchenko, Punnoose, Finkel'stein, and Klapwijk}}]{punnoose07}
\bibinfo{author}{\bibfnamefont{S.}~\bibnamefont{Anissimova}},
  \bibinfo{author}{\bibfnamefont{S.~V.} \bibnamefont{Kravchenko}},
  \bibinfo{author}{\bibfnamefont{A.}~\bibnamefont{Punnoose}},
  \bibinfo{author}{\bibfnamefont{A.~M.} \bibnamefont{Finkel'stein}},
  \bibnamefont{and} \bibinfo{author}{\bibfnamefont{T.~M.}
  \bibnamefont{Klapwijk}}, \bibinfo{journal}{Nature Physics}
  \textbf{\bibinfo{volume}{3}}, \bibinfo{pages}{707} (\bibinfo{year}{2007}).

\bibitem[{\citenamefont{Knyazev et~al.}(2007)\citenamefont{Knyazev,
  Omel'yanovskii, Pudalov, and Burmistrov}}]{pudalov_RG_JETP}
\bibinfo{author}{\bibfnamefont{D.~A.} \bibnamefont{Knyazev}},
  \bibinfo{author}{\bibfnamefont{O.~E.} \bibnamefont{Omel'yanovskii}},
  \bibinfo{author}{\bibfnamefont{V.~M.} \bibnamefont{Pudalov}},
  \bibnamefont{and} \bibinfo{author}{\bibfnamefont{I.~S.}
  \bibnamefont{Burmistrov}}, \bibinfo{journal}{JETP Lett.}
  \textbf{\bibinfo{volume}{84}}, \bibinfo{pages}{662} (\bibinfo{year}{2007}).

\bibitem[{\citenamefont{Altshuler and Aronov}(1985)}]{AAbook}
\bibinfo{author}{\bibfnamefont{B.~L.} \bibnamefont{Altshuler}}
  \bibnamefont{and} \bibinfo{author}{\bibfnamefont{A.~G.}
  \bibnamefont{Aronov}}, \emph{\bibinfo{title}{{Modern Problems in Condensed
  Matter Physics}}} (\bibinfo{publisher}{Elsevier, North Holland},
  \bibinfo{year}{1985}), chap. \bibinfo{chapter}{Electron-Electron Interactions
  in Disordered Systems}, p.~\bibinfo{pages}{1}.

\bibitem[{\citenamefont{Kawabata}(1981)}]{kawabata_zeeman}
\bibinfo{author}{\bibfnamefont{A.}~\bibnamefont{Kawabata}},
  \bibinfo{journal}{J. Phys. Soc. Jpn.} \textbf{\bibinfo{volume}{50}},
  \bibinfo{pages}{2461} (\bibinfo{year}{1981}).

\bibitem[{\citenamefont{Lee and Ramakrishnan}(1982)}]{tvr}
\bibinfo{author}{\bibfnamefont{P.~A.} \bibnamefont{Lee}} \bibnamefont{and}
  \bibinfo{author}{\bibfnamefont{T.~V.} \bibnamefont{Ramakrishnan}},
  \bibinfo{journal}{Phys. Rev. B} \textbf{\bibinfo{volume}{26}},
  \bibinfo{pages}{4009} (\bibinfo{year}{1982}).

\bibitem[{\citenamefont{Finkel'stein}(1983{\natexlab{a}})}]{sasha83b}
\bibinfo{author}{\bibfnamefont{A.~M.} \bibnamefont{Finkel'stein}},
  \bibinfo{journal}{JETP Lett.} \textbf{\bibinfo{volume}{37}},
  \bibinfo{pages}{517} (\bibinfo{year}{1983}{\natexlab{a}}).

\bibitem[{\citenamefont{Finkel'stein}(1984{\natexlab{a}})}]{sasha84b}
\bibinfo{author}{\bibfnamefont{A.~M.} \bibnamefont{Finkel'stein}},
  \bibinfo{journal}{Zh. Eksp. Teor. Fiz.} \textbf{\bibinfo{volume}{86}},
  \bibinfo{pages}{367} (\bibinfo{year}{1984}{\natexlab{a}}).

\bibitem[{\citenamefont{Castellani et~al.}(1984)\citenamefont{Castellani,
  Castro, Lee, and Ma}}]{pedagogical}
\bibinfo{author}{\bibfnamefont{C.}~\bibnamefont{Castellani}},
  \bibinfo{author}{\bibfnamefont{C.~D.} \bibnamefont{Castro}},
  \bibinfo{author}{\bibfnamefont{P.~A.} \bibnamefont{Lee}}, \bibnamefont{and}
  \bibinfo{author}{\bibfnamefont{M.}~\bibnamefont{Ma}}, \bibinfo{journal}{Phys.
  Rev. B} \textbf{\bibinfo{volume}{30}}, \bibinfo{pages}{527}
  (\bibinfo{year}{1984}).

\bibitem[{\citenamefont{Fukuyama}(1980)}]{fukuyama1}
\bibinfo{author}{\bibfnamefont{H.}~\bibnamefont{Fukuyama}},
  \bibinfo{journal}{J. Phys. Soc. Jpn.} \textbf{\bibinfo{volume}{49}},
  \bibinfo{pages}{649} (\bibinfo{year}{1980}).

\bibitem[{\citenamefont{Fukuyama}(1981)}]{fukuyama2}
\bibinfo{author}{\bibfnamefont{H.}~\bibnamefont{Fukuyama}},
  \bibinfo{journal}{J. Phys. Soc. Jpn.} \textbf{\bibinfo{volume}{50}},
  \bibinfo{pages}{3562} (\bibinfo{year}{1981}).

\bibitem[{\citenamefont{Finkel'stein}(1983{\natexlab{b}})}]{sasha83}
\bibinfo{author}{\bibfnamefont{A.~M.} \bibnamefont{Finkel'stein}},
  \bibinfo{journal}{Sov. Phys. JETP} \textbf{\bibinfo{volume}{57}},
  \bibinfo{pages}{97} (\bibinfo{year}{1983}{\natexlab{b}}).

\bibitem[{\citenamefont{Finkel'stein}(1990)}]{yellowbook}
\bibinfo{author}{\bibfnamefont{A.~M.} \bibnamefont{Finkel'stein}},
  \bibinfo{journal}{Sov. Sci. Rev. A, Phys. Rev.}
  \textbf{\bibinfo{volume}{14}}, \bibinfo{pages}{1} (\bibinfo{year}{1990}).

\bibitem[{\citenamefont{Raimondi et~al.}(1990)\citenamefont{Raimondi,
  Castellani, and Castro}}]{raimondi_zeeman}
\bibinfo{author}{\bibfnamefont{R.}~\bibnamefont{Raimondi}},
  \bibinfo{author}{\bibfnamefont{C.}~\bibnamefont{Castellani}},
  \bibnamefont{and} \bibinfo{author}{\bibfnamefont{C.~D.}
  \bibnamefont{Castro}}, \bibinfo{journal}{Phys. Rev. B}
  \textbf{\bibinfo{volume}{42}}, \bibinfo{pages}{4724} (\bibinfo{year}{1990}).

\bibitem[{\citenamefont{Castellani et~al.}(1998)\citenamefont{Castellani,
  Castro, and Lee}}]{castellani98}
\bibinfo{author}{\bibfnamefont{C.}~\bibnamefont{Castellani}},
  \bibinfo{author}{\bibfnamefont{C.~D.} \bibnamefont{Castro}},
  \bibnamefont{and} \bibinfo{author}{\bibfnamefont{P.~A.} \bibnamefont{Lee}},
  \bibinfo{journal}{Phys. Rev. B} \textbf{\bibinfo{volume}{57}},
  \bibinfo{pages}{R9381} (\bibinfo{year}{1998}).

\bibitem[{\citenamefont{Vitkalov et~al.}(2003)\citenamefont{Vitkalov, James,
  Narozhny, Sarachik, and Klapwijk}}]{ZNAfits_vitkalov}
\bibinfo{author}{\bibfnamefont{S.~A.} \bibnamefont{Vitkalov}},
  \bibinfo{author}{\bibfnamefont{K.}~\bibnamefont{James}},
  \bibinfo{author}{\bibfnamefont{B.~N.} \bibnamefont{Narozhny}},
  \bibinfo{author}{\bibfnamefont{M.~P.} \bibnamefont{Sarachik}},
  \bibnamefont{and} \bibinfo{author}{\bibfnamefont{T.~M.}
  \bibnamefont{Klapwijk}}, \bibinfo{journal}{Phys. Rev. B}
  \textbf{\bibinfo{volume}{67}}, \bibinfo{pages}{113310}
  (\bibinfo{year}{2003}).

\bibitem[{\citenamefont{Kuntsevich et~al.}(2007)\citenamefont{Kuntsevich,
  Klimov, Tarasenko, Averkiev, Pudalov, Kojima, and
  Gershenson}}]{tau_perp_gershenson}
\bibinfo{author}{\bibfnamefont{A.~Y.} \bibnamefont{Kuntsevich}},
  \bibinfo{author}{\bibfnamefont{N.~N.} \bibnamefont{Klimov}},
  \bibinfo{author}{\bibfnamefont{S.~A.} \bibnamefont{Tarasenko}},
  \bibinfo{author}{\bibfnamefont{N.~S.} \bibnamefont{Averkiev}},
  \bibinfo{author}{\bibfnamefont{V.~M.} \bibnamefont{Pudalov}},
  \bibinfo{author}{\bibfnamefont{H.}~\bibnamefont{Kojima}}, \bibnamefont{and}
  \bibinfo{author}{\bibfnamefont{M.~E.} \bibnamefont{Gershenson}},
  \bibinfo{journal}{Phys. Rev. B} \textbf{\bibinfo{volume}{75}},
  \bibinfo{pages}{195330} (\bibinfo{year}{2007}).

\bibitem[{\citenamefont{Burmistrov and
  Chtchelkatchev}(2008)}]{burmistrov_spinvalley}
\bibinfo{author}{\bibfnamefont{I.~S.} \bibnamefont{Burmistrov}}
  \bibnamefont{and} \bibinfo{author}{\bibfnamefont{N.~M.}
  \bibnamefont{Chtchelkatchev}}, \bibinfo{journal}{Phys. Rev. B}
  \textbf{\bibinfo{volume}{77}}, \bibinfo{pages}{195319}
  (\bibinfo{year}{2008}).

\bibitem[{\citenamefont{Punnoose}(2010)}]{punnoose_valley_RG}
\bibinfo{author}{\bibfnamefont{A.}~\bibnamefont{Punnoose}},
  \bibinfo{journal}{Phys. Rev. B} \textbf{\bibinfo{volume}{81}},
  \bibinfo{pages}{035306} (\bibinfo{year}{2010}).

\bibitem[{\citenamefont{Rahimi et~al.}(2003)\citenamefont{Rahimi, Anissimova,
  Sakr, Kravchenko, and Klapwijk}}]{dephasing_kravchenko}
\bibinfo{author}{\bibfnamefont{M.}~\bibnamefont{Rahimi}},
  \bibinfo{author}{\bibfnamefont{S.}~\bibnamefont{Anissimova}},
  \bibinfo{author}{\bibfnamefont{M.~R.} \bibnamefont{Sakr}},
  \bibinfo{author}{\bibfnamefont{S.~V.} \bibnamefont{Kravchenko}},
  \bibnamefont{and} \bibinfo{author}{\bibfnamefont{T.~M.}
  \bibnamefont{Klapwijk}}, \bibinfo{journal}{Phys. Rev. Lett.}
  \textbf{\bibinfo{volume}{91}}, \bibinfo{pages}{116402}
  (\bibinfo{year}{2003}).

\bibitem[{\citenamefont{Finkel'stein}(1984{\natexlab{b}})}]{sasha84}
\bibinfo{author}{\bibfnamefont{A.~M.} \bibnamefont{Finkel'stein}},
  \bibinfo{journal}{Z. Phys. B} \textbf{\bibinfo{volume}{56}},
  \bibinfo{pages}{189} (\bibinfo{year}{1984}{\natexlab{b}}).

\end{thebibliography}

\end{document}